# Thickness-Dependent Interfacial Coulomb Scattering in Atomically Thin Field-Effect Transistors

Song-Lin Li,[†,‡,*] Katsunori Wakabayashi,[†,*] Yong Xu,[†] Shu Nakaharai,[†] Katsuyoshi Komatsu,[†] Wen-Wu Li,[†] Yen-Fu Lin,[†] Alex Aparecido-Ferreira,[†] and Kazuhito Tsukagoshi[†,*]

[†]WPI Center for Materials Nanoarchitechtonics (WPI-MANA) and [‡]International Center for Young Scientist (ICYS), National Institute for Materials Science, Tsukuba, Ibaraki 305-0044, Japan

Email: li.songlin@nims.go.jp, wakabayashi.katsunori@nims.go.jp and tsukagoshi.kazuhito@nims.go.jp

**TOC graphic**

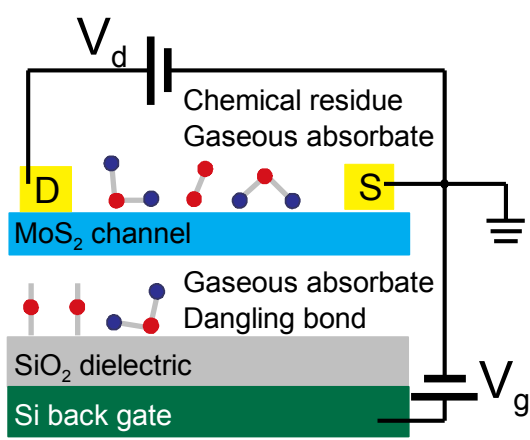
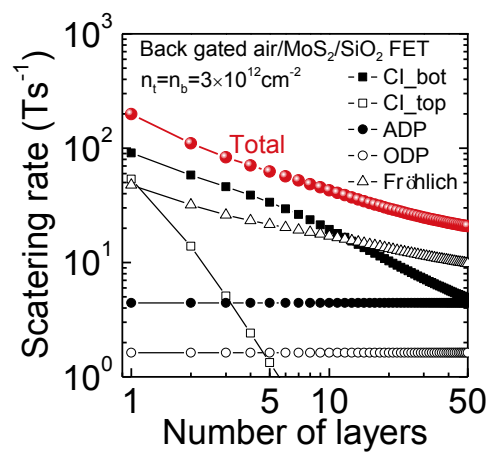

**Abstract:** Two-dimensional semiconductors are structurally ideal channel materials for the ultimate atomic electronics after silicon era. A long-standing puzzle is the low carrier mobility ($\mu$) in them as compared with corresponding bulk structures, which constitutes the main hurdle for realizing high-performance devices. To address this issue, we perform combined experimental and theoretical study on atomically thin $MoS_2$ field effect transistors with varying the number of $MoS_2$ layers (NLs). Experimentally, an intimate $\mu - \text{NL}$ relation is observed with a 10-fold degradation in $\mu$ for extremely thinned monolayer channels. To accurately describe the carrier scattering process and shed light on the origin of the thinning-induced mobility degradation, a generalized Coulomb scattering model is developed with strictly considering device configurative conditions, *i.e.*, asymmetric dielectric environments and lopsided carrier distribution. We reveal that the carrier scattering from interfacial Coulomb impurities (*e.g.*, chemical residues, gaseous adsorbates and surface dangling bonds) is greatly intensified in extremely thinned channels, resulting from shortened interaction distance between impurities and carriers. Such a pronounced factor may surpass lattice phonons and serve as dominant scatterers. This understanding offers new insight into the thickness induced scattering intensity, highlights the critical role of surface quality in electrical transport and would lead to rational performance improvement strategies for future atomic electronics.

Two-dimensional (2D) semiconductors are very attractive for the ultimate atomic field-effect transistor (FET) technology after silicon (Si) because of their unique dimensionality, thickness and flatness.[1–7] The intrinsic semiconducting nature, contrasting metallic graphene, allows for high on/off current ratios while the 2D structure offers compatibility to optical lithography and large scale fabrication. Besides, the ultrathin thickness enables aggressive device downscaling and high-density integration;[3–7] the atomic flatness makes them immune to surface roughness (SR) induced carrier scattering so that they can overcome the limitation of channel thickness ($t$) confronted by Si FETs.[8–10] Such merits lay the foundation for technologically viable atomic electronics. Recently, a wide spectrum of efforts has been made on the 2D MoS$_2$ flakes ranging from thickness characterization,[11–13] film growth,[14–17] to logic and optoelectronic devices.[18–21] In spite of the immunity to SR scattering, however, the ultrathin MoS$_2$ channels exhibit degraded electrical performance (carrier mobility $\mu \sim$ 1–10 cm$^2$V$^{-1}$s$^{-1}$)[1,22] with respect to corresponding bulk structures ($\mu \sim$ 200 cm$^2$V$^{-1}$s$^{-1}$)[23]. In an early report, Kis *et al.* advocated improving $\mu$ with dielectric screening and a greatly enhanced $\mu \sim$ 1000 cm$^2$V$^{-1}$s$^{-1}$ was reported,[2,24] but recently it was pointed out that the result was largely overestimated in the dual gate configuration.[25] Energy level matching between electrodes and channels was reported to be another way to improve $\mu$ in thick channels but seemed not so successful in extremely thinned atomic channels.[26] So far, the low performance in atomic FETs remains poorly understood and, without knowing the underlying mechanisms, it is difficult to develop effective strategies for realizing high-performance devices.

In this letter, we investigate the origin of electrical performance degradation in the planar FETs with ultrathin channels. A systematic evolution of $\mu$ with $t$ (from atomic to mesoscopic scale) is collected from back gated MoS$_2$/SiO$_2$ FETs. The $\mu - t$ relation is used as a unique clue to disentangle carrier scatterers with varied $t$ dependences (*i.e.*, SR, $t$-sensitive; Coulomb impurities, highly $t$-dependent; lattice phonons, moderately $t$-dependent; structural defects, $t$-independent) and to disclose the primary scattering mechanisms at atomic scale. To accurately describe the changes in carrier scattering process with device geometry, a generalized transport model is also developed with strictly considering real device configuration, *i.e.*, asymmetric dielectric environments and lopsided carrier distribution. It is revealed that lattice phonons dominates the scattering events in thick channels while interfacial Coulomb impurities (CIs) becomes the leading scatterers in extremely thinned atomic channels due to largely shortened interaction distance between CIs and carriers. It is tenable to explain the whole $\mu-t$ relation in terms of a CI and phonon coexisted transport model. Finally, the impact of each parameter on $\mu$ (*i.e.*, channel thickness $t$, density of CIs $n_i$, and gating geometry) is analyzed, for the purpose of developing rational strategies for $\mu$ improvement.

The experimental setup is as follows. The MoS$_2$ flakes, from 1 to 24 layers, were prepared by micromechanical cleavage from natural MoS$_2$ crystals and were transferred to oxygen plasma cleaned SiO$_2$/Si wafers (*p*-type, $\rho < 8\ \text{m}\Omega\ \text{cm}$). Figure 1a shows optical images for as-transferred large-area MoS$_2$ flakes with consecutive numbers of layers (NLs) from 2 to 6. The flake thickness is double checked by optical contrast[12,13] and interference Raman spectroscopy.[11] Here, the presence of multiple large flakes in local area offers several advantages. First, the close locations among flakes make it convenient to quickly identify NL by optical contrast. Besides, the consecutive NLs on the same substrate allow for a fair comparison of device performance because of nearly identical external scattering parameters. Furthermore, the large sizes (10–20 μm) enable us not only to etch channels with well defined width (*W*) and length (*L*) but also to deposit multiple electrodes to determine contact resistance ($R_c$) (Figure 1b). These procedures help to increase the experimental accuracy. Since the layer spacing of MoS$_2$ $c = 0.615$ nm,[27] the two thickness parameters (*t* and NL) are correlated by $t = c \times \text{NL}$. Most samples are within 16 layers (~10 nm) to suppress anisotropy induced c-axis access resistance ($R_a$), which causes $\mu$ decrease with increasing *t* when $t > 10$ nm.[26] All FETs are operated in back gate geometry with the highly doped Si as back gate and the 285 nm SiO$_2$ as gating dielectric (Figure 1c).

We systematically measured the dependence of electrical performance on NL. Figure 1d shows the conductivity ($\sigma = LI_d W^{-1} V_d^{-1}$) versus gate voltage ($V_g$) for NL from 1 to 5. The $\sigma$ exhibits positive dependence on NL, indicating higher $\mu$ in thicker samples. The $\sigma$ typically changes by 6–8 orders of magnitude, forming clear on and off current states (Inset of Figure 1d). The thin samples (NL = 1–3) normally show higher on/off ratio than thick ones (NL > 3), but their $\sigma$ curves are hard to reach the linear or sublinear regions at high $V_g$. They usually keep superlinear till the measurement limit of $V_g = 80$ V, indicating incomplete screening of external Coulomb scattering potentials. This feature is also reflected in the $\mu_{FE} - V_g$ plot (Figure 1e). Here the field-effect mobility is defined as the derivative of $\sigma$ to $V_g$, $\mu_{FE}(V_g) = C_{ox}^{-1} \partial \sigma / \partial V_g$ where $C_{ox} = 12$ nF cm$^{-2}$, which normally peaks at initial on state and decreases slightly with increasing $V_g$ in bulk FETs. No $\mu_{FE}$ peak is observed in the thin (NL = 1–3) samples up to $V_g = 80$ V while the peak occurs at 50–60 V for the thick 4- and 5-layer samples. The distinct $\mu$ saturation positions results from the varied scattering intensities due to $t$ variation because higher carrier density is required to fully screen the Coulomb scattering potential in thinned samples (See also Figure 4f).

In principle, the low work function of Ti would give rise to low contact barrier for the n-type MoS$_2$, but related devices often show performance degradation after long time storage, suggesting a degradation of contact interface with time. Thus, besides active Ti, stable Au electrodes are also used in separated devices. Consistent with previous work,[2,26] the Au-contacted devices show Ohmic contact to n-type MoS$_2$ after annealing, possibly due to the good wetting ability of Au. We find that the device stability of the Au-contacted samples is greatly enhanced as they can be stored for several months without noticeable degradation.

To avoid $\mu$ underestimation due to contact barrier, two techniques are employed to correct $\mu$, including the transfer length measurement (TLM)[28] and the Y function method (YFM).[29,30] The former relies on the extraction of a global $R_c$ and corrects $\mu$ by deducting the $R_c$ from total $R$ (Figure 1f), which requires multiple devices of varied $L$ and thus is valid only when large-area samples are available. The latter is more convenient than the former as it is applicable to individual devices. However, it is valid only for our thick samples because it bases on the linear fitting of a derived quantity Y function[29] ($YF = I_d g_m^{-1/2}$) in the sublinear regime of on state (Figure 1g). When possible, both methods are used for comparison. Most of the corrected $\mu$ values are consistent within 20% error between the two methods. Note that we only apply TLM for Au-contacted devices due to limited numbers of large-area MoS$_2$ flakes.

Figure 2 summarizes the evolution of $\mu$ with NL for the both Ti and Au contacts. To illustrate the contact effect, the raw and corrected values are independently plotted in Figure 2 a and b. The corrected values are typically enhanced by 20–50%, relative to raw data. The two types of electrodes exhibit a similar $\mu$ trend for thin samples from 1 to 10 layers, but show discrepancy for thick samples around 14 layers. The best $\mu$ of Ti-contacted 14-layer samples are 180 cm$^2$V$^{-1}$s$^{-1}$, consistent with the Sc-contacted ones.[26] The values are much higher than that in the Au-contacted 15-layer samples of $\mu \sim$ 60 cm$^2$V$^{-1}$s$^{-1}$. The origin of such a discrepancy of $\mu$ remains unclear at present and deserves a further investigation. Nevertheless, we adopt the highest observed rather than average $\mu$ values as intrinsic performance for later comparison with theoretical calculation. The main trend of $\mu$ versus NL is as below. The $\mu$ initially increases from ~20 to 180 cm$^2$V$^{-1}$s$^{-1}$ as NL changes from 1 to 14 and then decreases to ~30 cm$^2$V$^{-1}$s$^{-1}$ with further increasing NL due to the increase of c-axis access resistance $R_a$ at large NL values.[26] Unexpectedly, the $R_a$ cannot be effectively removed by the TLM or YFM correction from $\mu$ possibly because it is also highly $V_g$-dependent.[18] Here, we skip the discussion on the $R_a$-induced negative $\mu-t$ segment, which is out of motivation of this work, and only focus on the intrinsic positive $\mu-t$ behavior. We did not observe high $\mu$ in very thick samples (~50 nm) as reported in polymer separated samples by Bao *et al.*,[31] suggesting the critical role of the quality of channel/dielectric interface.

Typical scattering mechanisms in semiconductors include lattice phonons, CIs, SR, RIPs and structural defects. Despite the large numbers of candidates, the numbers of active primary sources responsible for the $\mu-t$ dependence are in fact limited. We first rule out the last one as the primary scatterer for the reason that the FETs of medium $t$ exhibit $\mu$ as high as 180 cm$^2$V$^{-1}$s$^{-1}$,[31] close to bulk samples. We also exclude RIP as the main mechanism because it is normally weak in low-permittivity (ε) SiO$_2$ gated FETs at low fields and only strong at high fields or upon high-ε gating.[32,33] The RIP scattering is expected to be weak in this study. We also tried fitting the global $\mu-t$ behavior with SR scattering (see Figure S5, Supporting Information), but found it is too $t$-sensitive to satisfy the low NL regime due to its sharp power law dependence ($\mu_{SR} \propto t^6$). Thus, we have to use lattice phonons and CIs to explain the device performance.

Two types of phonon scattering mechanisms are considered. One is lattice deformation potential (DP), arising from the lattice vibration induced local changes in lattice potential, which includes two sources: 1) quasi-elastic scattering on acoustic phonons (ADP) and 2) Inelastic scattering on optical phonons (ODP). The other is Fröhlich interaction, caused by the coupling of carriers to longitudinal optical phonon induced macroscopic polarization electric fields, which is present only in polar compounds. Since related studies have been well summarized in literature,[34–36] we will mainly focus on the CI interaction and consider the possibility of using phonons and CIs as principal scatterers to interpret the global $\mu-t$ behavior.

There are several limitations in previous CI scattering models.[37–41] No models simultaneously consider $t$ variation, lopsided carrier distribution and asymmetric dielectric environments around channels. For instance, simplifications of $t=\infty$ and $t=0$ are used for bulk Si FETs[37] (Figure 3a) and graphene[38,39] (Figure 3b), respectively. Besides, an exclusive carrier distribution (a pulse-like delta function) is used in graphene.[38,39] In superlattices[40,41] (Figure 3c), $t$ is tunable but symmetric carrier distributions (trigonometric) and dielectric environments ($\varepsilon_2 = \varepsilon_3$) are often employed. In striking contrast, the structure of air/MoS$_2$/SiO$_2$ represents a common dielectric/channel/dielectric trilayer system with finite $t$ (*e.g.*, $t$ spans from 1 to 10 layers; neither $t=0$ or $t=\infty$ simplification is valid.), asymmetric dielectric environments (*i.e.*, $\varepsilon_2 \neq \varepsilon_3$), and lopsided carrier distribution (close to gated dielectric). In fact, all the configurative parameters ($t$, carrier distribution, and dielectric environments) change the screening and polarization of carriers. Without considering the configurative differences, rigidly applying previous models on the common systems may cause large deviation. Thus, it is highly desired to develop a generalized model with accounting for realistic device configuration.

In our model, limitations are solved by removing all assumptions. We employ an electron wavefunction not only bridging atomic to mesoscopic scale but also accounting for the lopsided carrier distribution in channels. Besides, the effect of asymmetric dielectric environments is strictly considered. An effort was made on dealing with the infinite imaging charges in the two-boundary system with $\varepsilon_2 \neq \varepsilon_3$ (Figure S1, Supporting Information), which leads to complicated configurative form factors in scattering matrix elements and polarization function. The achievement of exact configurative form factors constitutes one of the main contributions of this work.

Strictly speaking, MoS$_2$ has a layered structure and is spatially discrete. In this study, we treated MoS$_2$ as a continuous medium. As will be seen later, such a simplification is reasonable and the calculation agrees with the experiment. We begin with an envelope electron wavefunction following the convention in bulk Si FETs,[37]

$$\phi(z) = \begin{cases} 0, & |z| > t/2 \\ (b^3/2)^{1/2}(z+t/2)e^{-b(z+t/2)/2}, & |z| \leq t/2 \end{cases} \quad (1)$$

where $z$ is the position in channel and $b$ is a variational parameter which depends on $t$ and $V_g$. The carrier distribution is expressed as $g(z) = |\phi(z)|^2$. In such a wavefunction, $b$ determines carrier distributions for different $t$ and $V_g$. Its accurate form should be derived from energy minimum principle and the final expression is expected to be rather complicated. For simplification, we assume $b = kV_g/t + b_{bulk}$ with $k$ a tunable coefficient in unit of V$^{-1}$. Such a form, though simple, is able to bridge $t$ in the whole range and correlate $V_g$ quickly, and thus well describes the dependence of carrier distribution on these two factors. It is easily justified that $b \to b_{bulk}(\infty)$ as $t \to \infty (0)$, representing the bulk ( or the pulse-like) limit. We found $k = 1/2$ is an appropriate value and is used in all calculation.

For electrically gated (including bulk) FETs, carriers are typically confined within nanometers range and behave as 2D electron gas.[37] Figure 4a shows the varied electron distributions at $V_g = 30$ V for three typical $t$ of 1, 5 and 14 layers. Arising from quantum effect and electrostatic equilibrium, the carriers, once gated, are attracted to the interface of gated dielectric and become lopsided for all $t$ values. For 14-layer channel, almost all carriers are confined within 5 layers. In contrast, the carriers in monolayer samples are "squeezed" into the monolayer and the average interaction distance ($d$) from carriers to interfacial CIs is largely reduced ($d < 1$ layer). The basic idea to use CI to interpret the strong $\mu$ degradation in monolayers is that the Coulomb scattering rate ($\tau$) increases with shortened $d$ because $\tau_{CI} \propto V(d)^2 \propto d^{-2}$ where $V(d)$ represents the Coulomb potential and is inversely proportional to $d$. For clarity, Figure 4b illustrates the trends of $d$ variation as NL is reduced from 5 to 1. The distances from the carrier distribution peak to the top and bottom surfaces are assigned as $d_t$ and $d_b$, respectively. As compared with the 5-layer sample, the monolayer sample has noticeably shortened $d_t$ and $d_b$ ($d_{t1} \ll d_{t5}$ and $d_{b1} \lessapprox d_{b5}$) and thus experiences stronger interfacial impurity scattering.

According to Boltzmann theory, the rate of elastic scattering in 2D systems is given by

$$\frac{1}{\tau_j(k)} = \frac{2D_0}{\hbar g_s g_v} \int_0^\pi \left|\frac{U_j(q)}{\varepsilon(q,T)}\right|^2 (1-\cos\theta) \mathrm{d}\theta \qquad (2)$$

where j index denotes different elastic scattering centers, $D_0$ the 2D density of states, $\hbar$ the reduced Planck constant, $g_s$ and $g_v$ the spin and valley degeneracy factors, $q = 2k\sin\frac{\theta}{2}$ the scattering vector with $\theta$ being the scattering angle, $T$ the temperature. All the configurative details are reflected in form factors and are included in the scattering matrix elements $U_j(q)$ and polarization function $\varepsilon(q,T)$. Related derivation and expressions are in Supporting Information.

The intensity of CI scattering is proportional to interfacial impurity density ($n_i$). By incorporating configurative form factors into $t$-dependent coefficients $\alpha_b(t)$ and $\alpha_t(t)$, the total $\tau$ can then be written as

$$\tau(t)^{-1} = \alpha_b(t)n_b + \alpha_t(t)n_t + \beta_F(t) + \beta_{ADP} + \beta_{ODP} \tag{3}$$

where $n_b$ and $n_t$ are the interfacial impurity densities at bottom and top channel surfaces, $\beta_F(t)$ the phonon contribution from Fröhlich interaction which is moderately $t$-dependent. $\beta_{ADP}$ and $\beta_{ODP}$ denote the DP contributions from quasi-elastic acoustic and inelastic optical phonons and they are $t$-independent. All of the coefficients are plotted in Figure S2, Supporting Information.

Note that the main sources of CIs are likely chemical residues, gaseous adsorbates and dielectric dangling bonds. The first two are supposed to attach to the top surface while the last two to the bottom. The CI densities are adopted with reference to the shift of threshold voltage ($V_t$) before and after long-time vacuum annealing (Figure S3, Supporting Information). Slightly higher values are used in consideration of possible incomplete CI removing by pure annealing. Figure 4c plots individual $\tau$ components for phonons[34–36] and CIs under assumption $n_t = n_b = 3 \times 10^{12} \, cm^{-2}$. The CI contribution from the gated (bottom) surface dominates within the calculated NL range from 1 to 10 layers. In contrast, the contribution from the ungated (top) surface is strong only for thin channels ($NL \leq 3$) and becomes weak or even negligible for thick channels. Such distinct dependences between the top and bottom CIs stems from different trends of $d_b$ and $d_t$, where $d_b$ slightly but $d_t$ considerably increases with NL. To further understand the roles of both channel surfaces, we considered varied $n_b/n_t$ ratios with fixing total CI density $n_i = n_t + n_b$ at $6 \times 10^{12} \, cm^{-2}$. For clarity, we directly translated $\tau$ into $\mu$ (Figure 4d). In general, $\mu$ increases monotonically with NL at each $n_b/n_t$ ratio, reflecting the critical role of interaction distance $d$. Besides, the higher the $n_b/n_t$ ratio, the lower $\mu$ is obtained. This trend is a natural result of the lopsided carrier distribution. In case of completely clean bottom surface $n_b = 0$, $\mu$ rapidly increases from 20 to 110 $cm^2V^{-1}s^{-1}$ as NL increases from 1 to 5, implying a direct solution to high $\mu$ by eliminating CIs on the surface of gated dielectric.

In Figure 4e, we carefully compare calculation with experiment which qualitatively agree. The experimental data are from not only our Ti- and Au-contacted FETs (blue square) but also the Sc-contacted low-barrier ones (red circle) by Das *et al.*[26] After correction, the trend of our data is consistent with theirs, indicating the effectiveness of the TLM and YFM techniques. The experimental $\mu$ data correspond to $n_b \sim 1-3\times10^{12}\,\text{cm}^{-2}$, indicating our samples contain large numbers of CIs, which is exactly the origin of low performance in the ultrathin devices.

Figure 4f shows the calculated dependence of $\mu$ on carrier density ($\propto V_g - V_t$) to understand its roles on scattering. There are two opposite effects with increasing carrier density. On the one hand, the screening ability of carriers to external Coulomb potential is enhanced, which elevates $\mu_{CI}$. On the other hand, the characteristic energy scale of carriers is pushed closer to that of polar LO at 48 meV, which increases the Fröhlich interaction and degrades $\mu_F$. The overall $\mu$ behavior is determined by these two effects *via* Matthiessen's rule. Thus, $\mu$ first increases and finally saturates with increasing $V_g - V_t$. In addition, the monolayer samples require higher $V_g - V_t$ to completely screen the external potential and thus it is unable to observe the $\mu$ saturation within limited $V_g - V_t$ regime, well explaining the observation in Figure 1e.

Next, we turn to the effect of dielectric environment and gate geometry. Detailed discussions remain elusive. Without knowing the basic theories, irrational experimental reports may arise.[2,19,24] In this sense, an appropriate theoretical insight is very important for directing experiment as well as developing rational strategies for high device performance. First, we investigate the screening effect of ungated dieelctric environment. $\mu$ increases at most twice in monolayers with the same level of $n_i$ but different top environments of air and HfO$_2$, comparing curve 2 with 1 in Figure 5. $\mu$ shows no enhancement for thick samples because the upper part of MoS$_2$ ($\varepsilon_{\text{MoS}_2} = 17.8 \sim \varepsilon_{\text{HfO}_2}$) itself serves as the high-ε environment.[31] Similarly, the influence of gate geometry is seen when comparing curves 2 and 4 where both $n_i$ and dielectric layers are same but gate geometry shifts from bottom- to top-gated. Under top-gating, the top high-ε dielectric starts to play dominant roles which leads to enhanced screening of impurity potentials. In this case, $\mu$ is enhanced by ~50% for all NL values. Third, we discuss the scheme of Kis *et al.* in which the above two strategies are employed. The results are given in curves 1 and 4. At most, the enhancement ratio reaches 3 (in monolayer), which is smaller than the change of average dielectric constant of environments $\frac{\varepsilon_2+\varepsilon_3'}{\varepsilon_2+\varepsilon_3} \approx 6$ ( $\varepsilon_2$ : $SiO_2$, $\varepsilon_3$ : air, $\varepsilon_3'$ : $HfO_2$ ) because of the involvement of screening-independent phonon contribution in scattering process. Thus, the dielectric screening can enhance $\mu$ to some extent, but is not the ultimate solution at all.

Finally, we discuss another essential but often neglected strategy to enhance $\mu$, *i.e.*, reducing impurity density $n_i$. In curve 5 (or 3), we adopt a low $n_i$ of $0.3 \times 10^{12} cm^{-2}$, a 10-fold cleaner surface. $\mu$ is greatly enhanced relative to high-$n_i$ devices in curve 4 (or 1). In our experiment, the oxygen plasma treatment reduces organic residues on $SiO_2$ surface but likely introduces large numbers of water molecules because they easily attach onto clean hydrophilic $SiO_2$ surface under ambient environment. In this sense, the surfacial water molecules may be one of the important sources as CIs. In view of this, we propose that the $\mu$ of 2D semiconductors would be greatly enhanced if the $n_i$ of bare $SiO_2$ surface can be largely reduced by post-annealing and performing transfer in a dry environment to avoid water absorption onto bottom surface. A hydrophobic dielectric surface would also reduce water absorption and help in the same way, which has been verified very recently by Fuhrer *et al.*[31] Simultaneously, chemical contaminants onto top surface of extremely thinned samples should also be minimized during device fabrication.

In conclusion, we provide an in-depth understanding on the carrier transport behavior for common trilayer structured FETs, especially on the dominant scattering mechanisms at atomic scale. To accurately capture the features of dielectric environment and carrier distribution, a generalized theoretical model for interfacial CI scattering is developed. Using channel thickness as a unique clue, we confirm the tenability of using CI and phonon coexisted models to explain the global dependence of $\mu$ on $t$ and uncover that the shortened Coulomb interaction distance is responsible for the degraded $\mu$ in ultrathin channels. As technological guidance, slightly thick channels, rather than extremely thinned monolayers, are more robust to extrinsic scattering and thus hold more favorable application promises. In addition, a clean interface is most crucial in realizing high-performance ultrathin-body FETs in atomic electronics.

**Supporting Information**

Derivation of CI scattering matrix elements, calculated scattering coefficients for CI and lattice phonon, estimation of CI density, effect of $V_g$ on $\mu$, and global fit with SR scattering. This material is available free of charge *via* the Internet at http://pubs.acs.org.

**Acknowledgements**

This work was partially supported by the FIRST Program from the JSPS.

**References**

1. Novoselov, K. S.; Jiang, D.; Schedin, F.; Booth, T. J.; Khotkevich, V. V.; Morozov, S. V.; Geim, A. K., Two-Dimensional Atomic Crystals. *Proc. Natl. Acad. Sci. USA* **2005**, *102*, 10451–10453.
2. Radisavljevic, B.; Radenovic, A.; Brivio, J.; Giacometti, V.; Kis, A., Single-Layer MoS$_2$ Transistors. *Nat. Nanotechnol.* **2011**, *6*, 147–150.
3. Schwierz, F., Graphene Transistors. *Nat. Nanotechnol.* **2010**, *5*, 487–496.
4. Wong, H. S. P., Beyond the Conventional Transistor. *IBM J. Res. Dev.* **2002**, *46*, 133–168.
5. Ieong, M.; Doris, B.; Kedzierski, J.; Rim, K.; Yang, M., Silicon Device Scaling to the Sub-10-nm Regime. *Science* **2004**, *306*, 2057–2060.
6. Vogel, E. M., Technology and Metrology of New Electronic Materials and Devices. *Nat. Nanotechnol.* **2007**, *2*, 25–32.
7. Frank, D.; Taur, Y.; Wong, H.-S., Generalized Scale Length for Two-Dimensional Effects in MOSFETs. *IEEE Electron Device Lett.* **1998**, *19*, 385–387.
8. Sakaki, H.; Noda, T.; Hirakawa, K.; Tanaka, M.; Matsusue, T., Interface Roughness Scattering in GaAs/AlAs Quantum Wells. *Appl. Phys. Lett.* **1987**, *51*, 1934–1936.
9. Jin, S.; Fischetti, M.; Tang, T.-W., Modeling of Surface-Roughness Scattering in Ultrathin-Body SOI MOSFETs. *IEEE Trans. Electron Devices* **2007**, *54*, 2191–2203.
10. Gomez, L.; Aberg, I.; Hoyt, J. L., Electron Transport in Strained-Silicon Directly on Insulator Ultrathin-Body N-MOSFETs with Body Thickness Ranging from 2 to 25 nm. *IEEE Electron Device Lett.* **2007**, *28*, 285–287.
11. Li, S.-L.; Miyazaki, H.; Song, H.; Kuramochi, H.; Nakaharai, S.; Tsukagoshi, K., Quantitative Raman Spectra and Reliable Thickness Identification for Atomic Layers on Insulating Substrates. *ACS Nano* **2012**, *6*, 7381–7388.
12. Castellanos-Gomez, A.; Agraït, N.; Rubio-Bollinger, G., Optical Identification of Atomically Thin Dichalcogenide Crystals. *Appl. Phys. Lett.* **2010**, *96*, 213116.
13. Late, D. J.; Liu, B.; Matte, H. S. S. R.; Rao, C. N. R.; Dravid, V. P., Rapid Characterization of Ultrathin Layers of Chalcogenides on SiO$_2$/Si Substrates. *Adv. Funct. Mater.* **2012**, *22*, 1894–1905.

14. Liu, K.-K.; Zhang, W.; Lee, Y.-H.; Lin, Y.-C.; Chang, M.-T.; Su, C.; Chang, C.-S.; Li, H.; Shi, Y.; Zhang, H.; et al., Growth of Large-Area and Highly Crystalline MoS$_2$ Thin Layers on Insulating Substrates. *Nano Lett.* **2012**, *12*, 1538–1544.
15. Shi, Y.; Zhou, W.; Lu, A.-Y.; Fang, W.; Lee, Y.-H.; Hsu, A. L.; Kim, S. M.; Kim, K. K.; Yang, H. Y.; Li, L.-J.; et al., van der Waals Epitaxy of MoS$_2$ Layers Using Graphene as Growth Templates. *Nano Lett.* **2012**, *12*, 2784–2791.
16. Zhan, Y.; Liu, Z.; Najmaei, S.; Ajayan, P. M.; Lou, J., Large-Area Vapor-Phase Growth and Characterization of MoS$_2$ Atomic Layers on a SiO$_2$ Substrate. *Small* **2012**, *8*, 966–971.
17. Lee, Y.-H.; Zhang, X.-Q.; Zhang, W.; Chang, M.-T.; Lin, C.-T.; Chang, K.-D.; Yu, Y.-C.; Wang, J. T.-W.; Chang, C.-S.; Li, L.-J.; et al., Synthesis of Large-Area MoS$_2$ Atomic Layers with Chemical Vapor Deposition. *Adv. Mater.* **2012**, *24*, 2320–2325.
18. Yu, W. J.; Li, Z.; Zhou, H.; Chen, Y.; Wang, Y.; Huang, Y.; Duan, X., Vertically Stacked Multi-Heterostructures of Layered Materials for Logic Transistors and Complementary Inverters. *Nat. Mater.* **2013**, *12*, 246–252.
19. Radisavljevic, B.; Whitwick, M. B.; Kis, A., Integrated Circuits and Logic Operations Based on Single-Layer MoS$_2$. *ACS Nano* **2011**, *5*, 9934–9938.
20. Wang, H.; Yu, L.; Lee, Y.-H.; Shi, Y.; Hsu, A.; Chin, M. L.; Li, L.-J.; Dubey, M.; Kong, J.; Palacios, T., Integrated Circuits Based on Bilayer MoS$_2$ Transistors. *Nano Lett.* **2012**, *12*, 4674–4680.
21. Lee, H. S.; Min, S.-W.; Chang, Y.-G.; Park, M. K.; Nam, T.; Kim, H.; Kim, J. H.; Ryu, S.; Im, S., MoS$_2$ Nanosheet Phototransistors with Thickness-Modulated Optical Energy Gap. *Nano Lett.* **2012**, *12*, 3695–3700.
22. Ghatak, S.; Pal, A. N.; Ghosh, A., Nature of Electronic States in Atomically Thin MoS$_2$ Field-Effect Transistors. *ACS Nano* **2011**, *5*, 7707–7712.
23. Fivaz, R.; Mooser, E., Mobility of Charge Carriers in Semiconducting Layer Structures. *Phys. Rev.* **1967**, *163*, 743–755.
24. Lembke, D.; Kis, A., Breakdown of High-Performance Monolayer MoS$_2$ Transistors. *ACS Nano* **2012**, *6*, 10070–10075.
25. Fuhrer, M. S.; Hone, J., Measurement of Mobility in Dual-Gated MoS$_2$ Transistors. *Nat. Nanotechnol.* **2013**, *8*, 146–147.
26. Das, S.; Chen, H.-Y.; Penumatcha, A. V.; Appenzeller, J., High Performance Multilayer MoS$_2$ Transistors with Scandium Contacts. *Nano Lett.* **2013**, *13*, 100–105.
27. Wilson, J. A.; Yoffe, A. D., The Transition Metal Dichalcogenides: Discussion and Interpretation of Observed Optical, Electrical and Structural Properties. *Adv. Phys.* **1969**, *18*, 193–335.
28. Zaumseil, J.; Baldwin, K. W.; Rogers, J. A., Contact Resistance in Organic Transistors that Use Source and Drain Electrodes Formed by Soft Contact Lamination. *J. Appl. Phys.* **2003**, *93*, 6117–6124.
29. Ghibaudo, G., New Method for the Extraction of MOSFET Parameters. *Electron. Lett.* **1988**, *24*, 543–545.
30. Xu, Y.; Minari, T.; Tsukagoshi, K.; Chroboczek, J. A.; Ghibaudo, G., Direct Evaluation of Low-Field Mobility and Access Resistance in Pentacene Field-Effect Transistors. *J. Appl. Phys.* **2010**, *107*, 114507.
31. Bao, W.; Cai, X.; Kim, D.; Sridhara, K.; Fuhrer, M. S., High Mobility Ambipolar MoS$_2$ Field-Effect Transistors: Substrate and Dielectric Effects. *Appl. Phys. Lett.* **2013**, *102*, 042104.
32. Moore, B. T.; Ferry, D. K., Remote Polar Phonon Scattering in Si Inversion Layers. *J. Appl. Phys.* **1980**, *51*, 2603–2605.
33. Fischetti, M. V.; Neumayer, D. A.; Cartier, E. A., Effective Electron Mobility in Si Inversion Layers in Metal-Oxide-Semiconductor Systems with a High-$\kappa$ Insulator: The Role of Remote Phonon Scattering. *J. Appl. Phys.* **2001**, *90*, 4587–4608.

34. Kaasbjerg, K.; Thygesen, K. S.; Jacobsen, K. W., Phonon-Limited Mobility in n-Type Single-Layer MoS$_2$ from First Principles. *Phys. Rev. B* **2012**, *85*, 115317.
35. Gelmont, B. L.; Shur, M.; Stroscio, M., Polar Optical-Phonon Scattering in Three- and Two-Ddimensional Electron Gases. *J. Appl. Phys.* **1995**, *77*, 657–660.
36. Kim, S.; Konar, A.; Hwang, W.-S.; Lee, J. H.; Lee, J.; Yang, J.; Jung, C.; Kim, H.; Yoo, J.-B.; Choi, J.-Y.; *et al.*, High-Mobility and Low-Power Thin-Film Transistors Based on Multilayer MoS$_2$ Crystals. *Nat. Commun.* **2012**, *3*, 1011.
37. Ando, T.; Fowler, A. B.; Stern, F., Electronic Properties of Two-Dimensional Systems. *Rev. Mod. Phys.* **1982**, *54*, 437–672.
38. Ando, T., Screening Effect and Impurity Scattering in Monolayer Graphene. *J. Phys. Soc. Jpn.* **2006**, *75*, 074716.
39. Das Sarma, S.; Adam, S.; Hwang, E. H.; Rossi, E., Electronic Transport in Two-Dimensional Graphene. *Rev. Mod. Phys.* **2011**, *83*, 407–470.
40. Gold, A., Electronic Transport Properties of a Two-Dimensional Electron Gas in a Silicon Quantum-Well Structure at Low Temperature. *Phys. Rev. B* **1987**, *35*, 723–733.
41. Jena, D.; Konar, A., Enhancement of Carrier Mobility in Semiconductor Nanostructures by Dielectric Engineering. *Phys. Rev. Lett.* **2007**, *98*, 136805.

**Figures and captions**

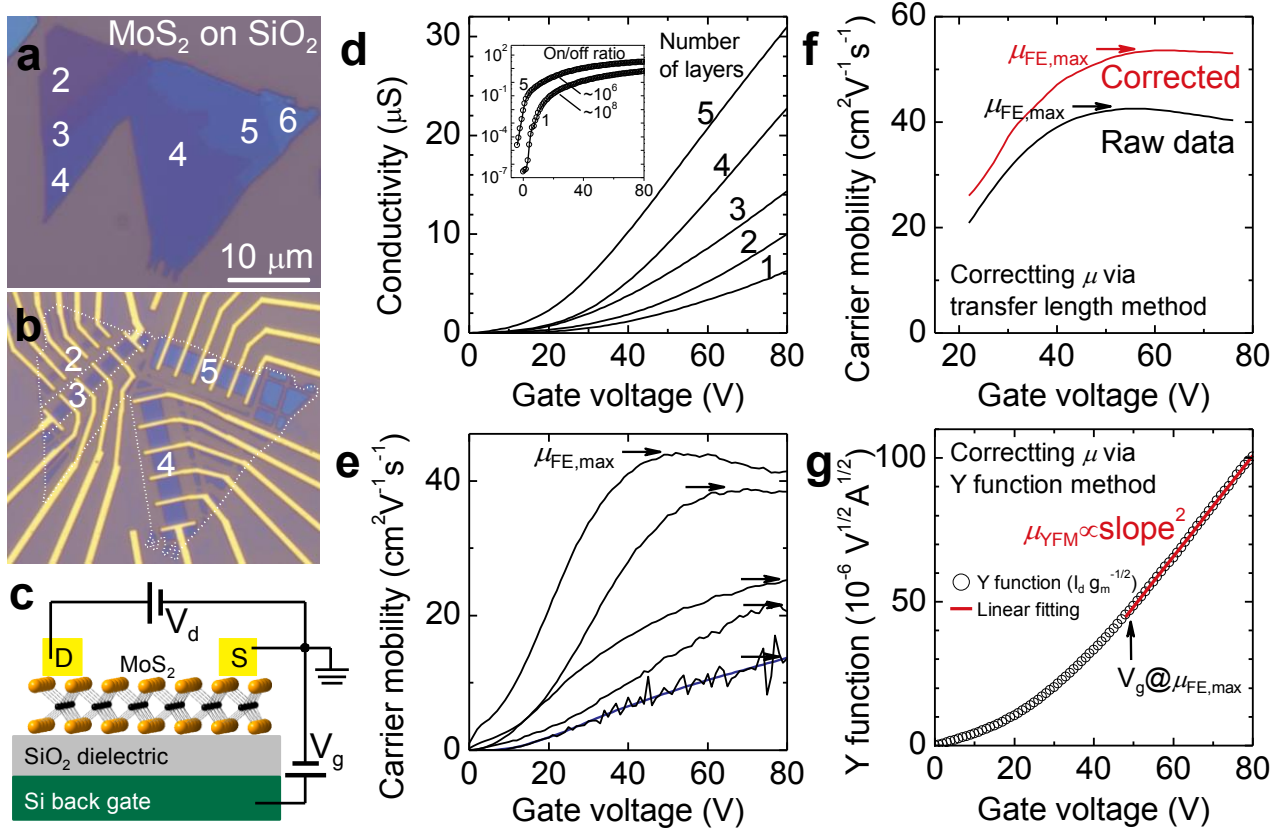

**Figure 1** (a) and (b) Optical images for as-transferred $MoS_2$ flakes with consecutive numbers of layers from 2 to 6 and corresponding FET devices. (c) Schematic diagram for a back-gated $MoS_2$ FET. (d) and (e) NL-dependent electrical performance (conductivity and carrier mobility) with FET channels changing from 1 to 5 layers. The inset in (d) is the corresponding logarithmic plot to show the on/off current ratio. (f) Mobility curves before and after TLM correction. (g) Applying linear fitting on Y function to extract $\mu_{YFM}$.

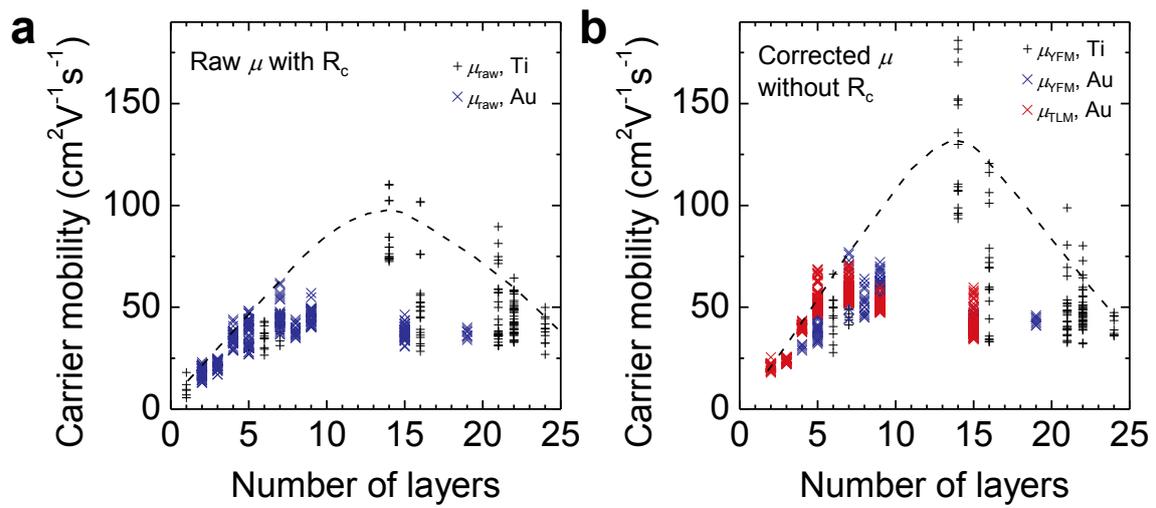

**Figure 2** Summary of carrier mobility $\mu$ as a function of channel thickness. (a) Raw and (b) corrected $\mu$ values for both Ti- and Au-contacted FETs. The dashed lines are guided for eyes.

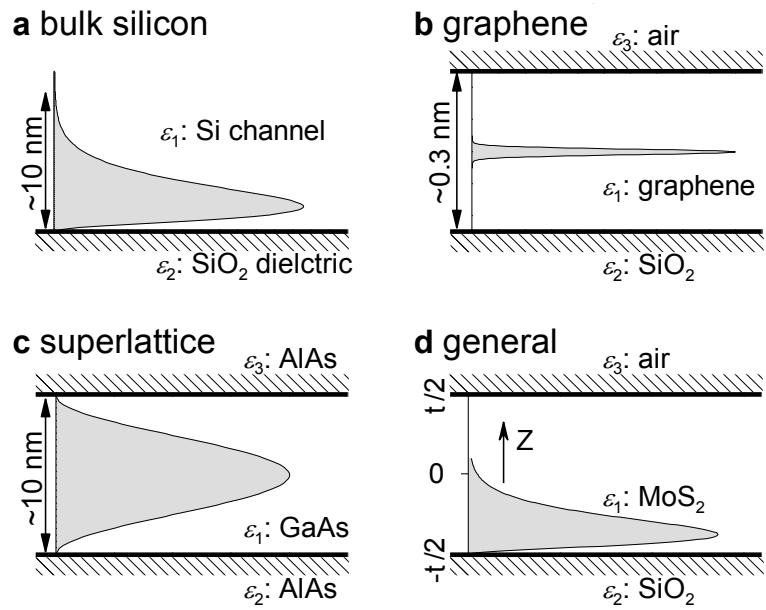

**Figure 3** Schematic diagrams of dielectric environments and carrier distributions for different device configurations. (a) Bulk silicon: one boundary which produces only one image charge.[37] (b) Graphene: negligible $t$ for the middle layer and a simple pulse-like carrier distribution.[38,39] (c) Superlattice: symmetric dielectrics and trigonometric wavefunction.[40,41] (d) A common channel: 1) two boundaries which produce infinite image charges; 2) a lopsided carrier distribution which leads to complicated configurative form factors in scattering matrix elements and dielectric function.

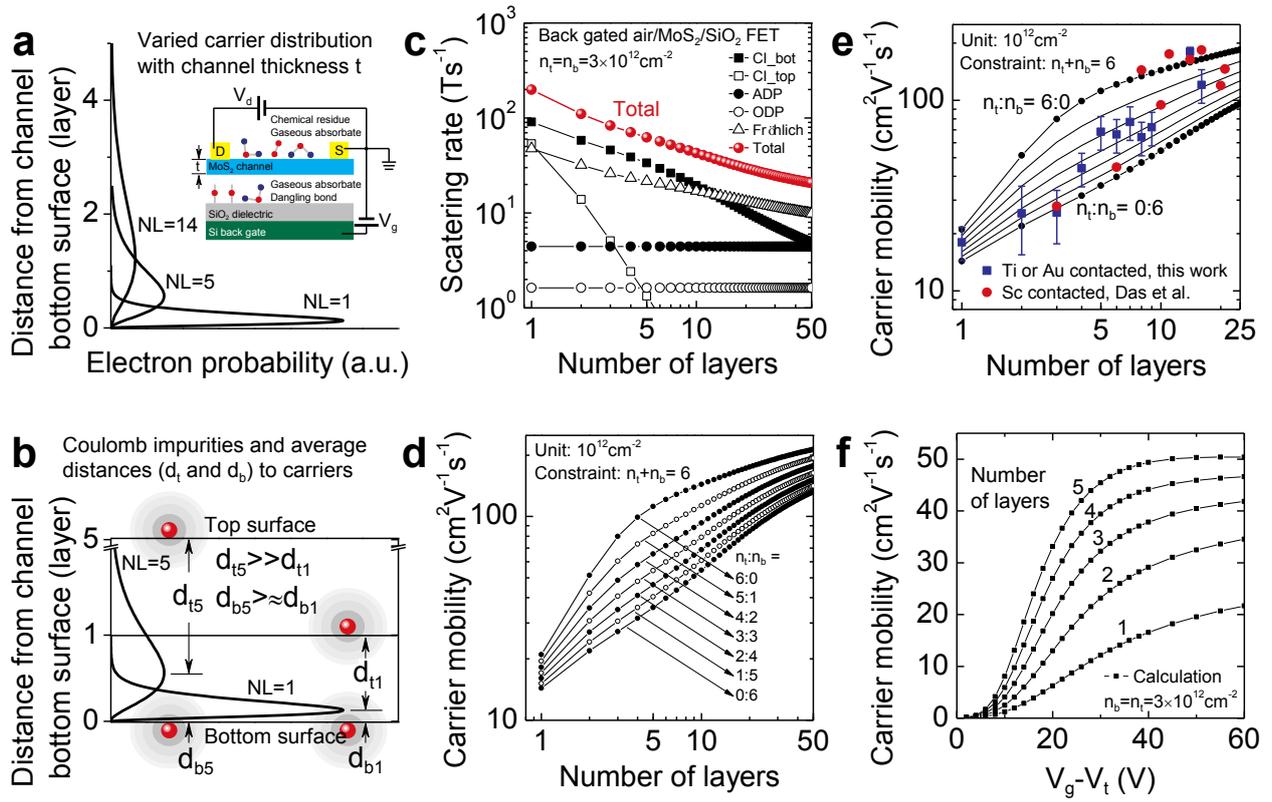

**Figure 4** (a) Carrier distributions in 1-, 5- and 14-layer channels. Inset: Diagram of charged impurities (*e.g.*, chemical residues, gaseous adsorbates and surface dangling bonds) located on the top and bottom channel surfaces, which are the leading scatterers in ultrathin channels. (b) Comparison of interaction distances $d_t$ and $d_b$ between 1- and 5-layer channels. The red dots and circular shades denote the interfacial charged impurities and corresponding scattering potential. (c) Calculated scattering rates for a back-gated air/MoS$_2$/SiO$_2$ structure assuming $n_t = n_b = 3\times10^{12}\text{cm}^{-2}$. (d) Calculated carrier mobilities by tuning the $n_t / n_b$ ratio under a total $n_i = n_t + n_b = 6\times10^{12}\text{cm}^{-2}$. (e) Comparison between calculation and experiment. (f) Calculated dependence of $\mu$ on carrier density ($\propto V_g - V_t$).

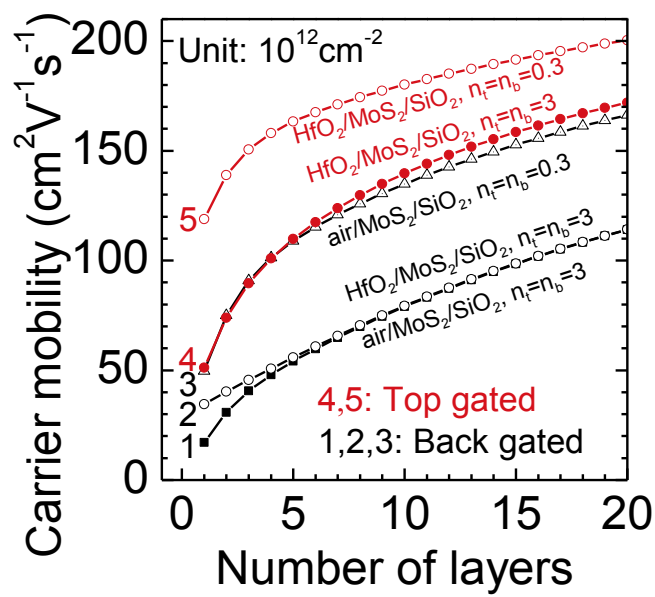

**Figure 5** Effect of experimental conditions (high-ε environment, gate geometry and impurity density) on carrier mobility.

SUPPORTING INFORMATION

**Thickness-Dependent Interfacial Coulomb Scattering in Atomically Thin Field-Effect Transistors**

Song-Lin Li,[†,‡,*] Katsunori Wakabayashi,[†,*] Yong Xu,[†] Shu Nakaharai,[†] Katsuyoshi Komatsu,[†] Wen-Wu Li,[†] Yen-Fu Lin,[†] Alex Aparecido-Ferreira,[†] and Kazuhito Tsukagoshi[†,*]

[†]WPI Center for Materials Nanoarchitechtonics (WPI-MANA) and [‡]International Center for Young Scientist (ICYS), National Institute for Materials Science, Tsukuba, Ibaraki 305-0044, Japan

E-mail: li.songlin@nims.go.jp, wakabayashi.katsunori@nims.go.jp and tsukagoshi.kazuhito@nims.go.jp

**Outline**

1. **Derivation of CI scattering matrix elements**
2. **Calculated scattering coefficients for CI and lattice phonon**
3. **Estimation of CI density**
4. **Effect of $V_g$ on $\mu$**
5. **Global fit with SR scattering**

1. Derivation of CI scattering matrix elements

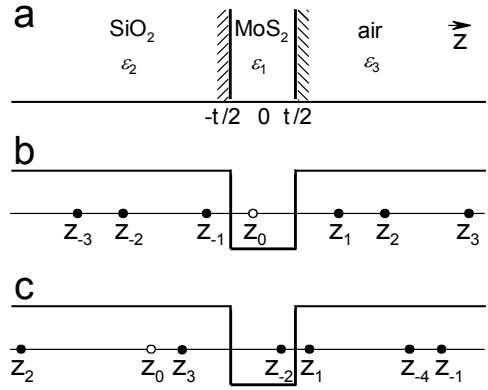

**Figure S1** (a) The dielectric environements for a common trilayer structure, where the semiconductor channel is sandwiched by two asymmetric dielectrics. (b) and (c) Applying the image method to derive the Coulomb force in the trilayer structure where the two boundaries make infinite charge images. The Coulomb force for two point charges located in the central layer is considered in (b). One charge is fixed and the other charge is mirrored through the two dielectric boundaries. This mirroring process produces infinite charge images at position $z_n = nt + (-1)^n z_0$, n=0, ±1, ±2 ... (c) plots the positions of image charges fo the case when one charge is located in the center and the other in the left dielctric.

For a common trilayer FET with asymmetric dielectric environments (Figure S1a), the expressions of Coulomb forces between two point charges at any positions (3D coordinates $\vec{\rho}_a$ and $\vec{\rho}_b$ where $\vec{\rho} = (\vec{r}, z)$ and $\vec{r}$ denotes the 2D coordinate in the XY plane) are listed in Reference S1. The Coulomb force is achieved by the method of image charges. To calculate the Coulomb force between two point charges, one charge is often fixed and the other one is mirror imaged by the two boundaries. This imaging process produces infinite image charges because the one boundary always produces a new image to that from the other. Thus, the final expression of the Coulomb force is an infinite series.

In our derivation, we need considering the following two cases: 1) both charges are located in the center (Figure S1b), used for deriving the dielectric polarization function $\varepsilon(q,T)$ which dealing the interaction and polarization of carriers; 2) one is located in center and the other in right (Figure S1c), for deriving the scattering matrix elements $U_j(q)$ which consider the interaction of an external charged impurity with carriers.

For case 1, the force is given by

$$F_{CC}(\vec{\rho}_a,\vec{\rho}_b) = \sum_{n=-\infty}^{\infty} \frac{e^2 \xi^{|n|}}{\varepsilon_1[(\vec{r}_a-\vec{r}_b)^2+(z_a-z_b-2nt)^2]^{1/2}} + \frac{\varepsilon_1-\varepsilon_3}{\varepsilon_1+\varepsilon_3}\sum_{n=0}^{\infty}\frac{e^2\xi^n}{\varepsilon_1[(\vec{r}_a-\vec{r}_b)^2+(z_a+z_b-(2n+1)t)^2]^{1/2}}$$
$$+\frac{\varepsilon_1-\varepsilon_3}{\varepsilon_1+\varepsilon_3}\sum_{n=0}^{\infty}\frac{e^2\xi^n}{\varepsilon_1[(\vec{r}_a-\vec{r}_b)^2+(z_a+z_b+(2n+1)t)^2]^{1/2}} \tag{S1}$$

where $\xi = \frac{\varepsilon_1-\varepsilon_2}{\varepsilon_1+\varepsilon_2}\frac{\varepsilon_1-\varepsilon_3}{\varepsilon_1+\varepsilon_3}$. For case 2, the force is given by

$$F_{LC}(\vec{\rho}_a,\vec{\rho}_b) = \frac{2}{\varepsilon_1+\varepsilon_2}\sum_{n=0}^{\infty}\frac{e^2\xi^n}{[(\vec{r}_a-\vec{r}_b)^2+(z_a-z_b-2nt)^2]^{1/2}}$$
$$+\frac{2}{\varepsilon_1+\varepsilon_2}\frac{\varepsilon_1-\varepsilon_3}{\varepsilon_1+\varepsilon_3}\sum_{n=0}^{\infty}\frac{e^2\xi^n}{[(\vec{r}_a-\vec{r}_b)^2+(z_a+z_b-(2n+1)t)^2]^{1/2}} \tag{S2}$$

With random phase approximation the polarization function $\varepsilon(q,T)$ can be derived from

$$\varepsilon(q,T) = 1 + \frac{e^2\Pi(q,T)}{2\varepsilon_0\varepsilon_1 q}\int_{-t/2}^{t/2}\int_{-t/2}^{t/2} g(z_a,b)g(z_b,b)\text{Fr}[F_{CC}(\vec{\rho}_a,\vec{\rho}_b)]dz_a dz_b \tag{S3}$$

where $e$ is the elementary charge, $\Pi(q,T)$ is the 2D finite-temperature electron polarizability, $\varepsilon_0$ is the vacuum permittivity, $\varepsilon_i$ (i = 1, 2, 3) are the relative dielectric constants for related layers. Fr[ ] denotes the 2D Fourier transformation from real to momentum space. Substituting Eq. S1 into S3, carefully dealing with the summations and merging the terms, one can finally obtain

$$\varepsilon(q,T) = 1 + \frac{e^2\Pi(q,T)}{2\varepsilon_0\varepsilon_1 q}\left(F_{ee0}(q,t) + \frac{2\frac{\varepsilon_1-\varepsilon_2}{\varepsilon_1+\varepsilon_2}\frac{\varepsilon_1-\varepsilon_3}{\varepsilon_1+\varepsilon_3}F_{ee1}(q,t)+\frac{\varepsilon_1-\varepsilon_3}{\varepsilon_1+\varepsilon_3}F_{ee2}(q,t)+\frac{\varepsilon_1-\varepsilon_2}{\varepsilon_1+\varepsilon_2}F_{ee3}(q,t)}{1-\frac{\varepsilon_1-\varepsilon_2}{\varepsilon_1+\varepsilon_2}\frac{\varepsilon_1-\varepsilon_3}{\varepsilon_1+\varepsilon_3}e^{-2qt}}\right) \tag{S4}$$

where $F_{eei}(q,t)$ (i = 0,1,2,3) are the configurative form factors originating from asymmetric environments.[S1] Detailed expressions are

$$F_{ee0}(q,t) = \frac{b(8b^2+9bq+3q^2)}{8(b+q)^3}$$

$$F_{ee1}(q,t) = \frac{b^6 e^{-2t(b+q)}(2-2e^{t(b-q)}+t(b-q)(2+t(b-q)))(2-2e^{t(b+q)}+t(b+q)(2+t(b+q)))}{4(b^2-q^2)^3}$$

$$F_{ee2}(q,t) = \frac{b^6 e^{-2bt}(2-2e^{t(b-q)}+t(b-q)(2+t(b-q)))^2}{4(b-q)^6}$$

$$F_{ee3}(q,t) = \frac{b^6 e^{-2t(b+q)}(-2+2e^{t(b+q)}-t(b+q)(2+t(b+q)))^2}{4(b+q)^6}.$$

When the right dielectric has a same dielectric constant with the channel ($\varepsilon_3 = \varepsilon_1$), Equation 3 is simplified into a two-layer system, exactly same as Ando's model for bulk Si FETs.[37]

We also managed to reach analytical expressions for the different scattering matrix elements for CIs located in the gated and ungated dielectric environments. For planar CIs in the bottom and top dielectrics with positions $z_b$ and $z_t$ and surface densities $n_b$ and $n_t$, the scattering matrix elements $U_b(q, T)$ can be derived from

$$U_{ib}(q,b,t,z_b) = n_b^{1/2} \frac{e^2 b^3 e^{qz_b}}{4\varepsilon_0 \varepsilon_1 q} \int_{-t/2}^{t/2} g(z_a,b) \operatorname{Fr}[F_{LC}(\vec{\rho}_a, \vec{\rho}_b)] dz_a \qquad z_b \leq -t/2 \qquad (S5)$$

with the final form

$$U_{ib}(q,b,t,z_b) = n_b^{1/2} \frac{e^2 b^3 e^{qz_b}}{4\varepsilon_0 \varepsilon_1 q} \frac{F_{ib1}(q,b,t) + \frac{\varepsilon_1-\varepsilon_3}{\varepsilon_1+\varepsilon_3} F_{ib2}(q,b,t)}{(1 - \frac{\varepsilon_1-\varepsilon_2}{\varepsilon_1+\varepsilon_2} \frac{\varepsilon_1-\varepsilon_3}{\varepsilon_1+\varepsilon_3} e^{-2qt}) \frac{\varepsilon_1+\varepsilon_2}{2\varepsilon_1}} \qquad z_b \leq -t/2 \qquad (S6)$$

where the form factors $F_{iX1}(q,b,t)$ and $F_{iX2}(q,b,t)$ (X = d or t) arise from the presence of a point charge and its first-order image charge, and have the forms below

$$F_{ib1}(q,b,t) = -\frac{b^3 e^{-t(b+q)}(2 - 2e^{t(b+q)} + t(b+q)(2 + t(b+q)))}{2(b+q)^3}$$

$$F_{ib2}(q,b,t) = -\frac{b^3 e^{-t(b+q)}(2 - 2e^{t(b-q)} + t(b-q)(2 + t(b-q)))}{2(b-q)^3}$$

Again for verification, Equation 4 (or 5) evolves into Ando's model if taking $\varepsilon_3 = \varepsilon_1$ (or $\varepsilon_2 = \varepsilon_1$).

Similarly, one can easily obtain the expression for $U_{te}(q, T)$ as below

$$U_{it}(q,b,t,z_t) = n_t^{1/2} \frac{e^2 b^3 e^{qz_t}}{4\varepsilon_0 \varepsilon_1 q} \frac{F_{it1}(q,b,t) + \frac{\varepsilon_1-\varepsilon_2}{\varepsilon_1+\varepsilon_2} F_{it2}(q,b,t)}{(1 - \frac{\varepsilon_1-\varepsilon_2}{\varepsilon_1+\varepsilon_2} \frac{\varepsilon_1-\varepsilon_3}{\varepsilon_1+\varepsilon_3} e^{-2qt}) \frac{\varepsilon_1+\varepsilon_3}{2\varepsilon_1}} \qquad z_t \geq t/2 \qquad (S7)$$

with the form factors

$$F_{it1}(q,b,t) = -\frac{b^3 e^{-bt}(2 - 2e^{t(b-q)} + t(b-q)(2 + t(b-q)))}{2(b-q)^3}$$

$$F_{it2}(q,b,t) = -\frac{b^3 e^{-t(b+2q)}(2 - 2e^{t(b+q)} + t(b+q)(2 + t(b+q)))}{2(b+q)^3}$$

2. Calculated scattering coefficients for CI and lattice phonon

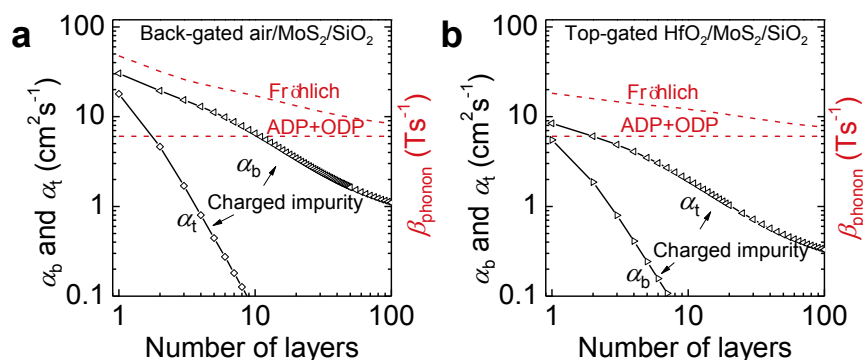

**Figure S2** The values of CI and phonon scattering coefficients $\alpha_b$, $\alpha_t$ and $\beta_{phonon}$ for (a) back-gated air/MoS$_2$/SiO$_2$ and (b) top-gated HfO$_2$/MoS$_2$/SiO$_2$ FETs.

3. Estimation of CI density

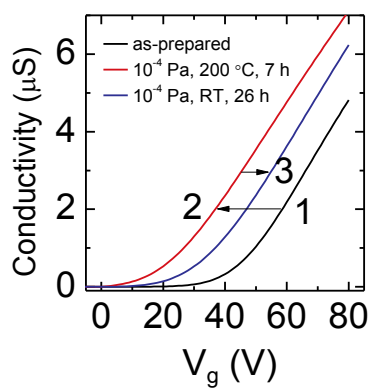

**Figure S3** Shifts of threshold voltage ($V_t$) under varied environments: 1) as-prepared, 2) vacuum annealing at 200 °C for 7 hours, and 3) 10$^{-4}$ Pa, room temperature for 26 hours.

The Coulomb impurities at the bottom surface may arise from water molecules and dangling bonds on SiO$_2$ and the impurities at top surface could stem from gaseous adsorbates as well as chemical residues introduced during device fabrication. It proved difficult to clean these impurities completely on the fragile MoS$_2$ by simple annealing which tends to decompose or be oxidized at elevated temperature. We have ever carefully checked the role of long-time vacuum annealing (10$^{-4}$ Pa, 200 °C for 7 hours) but found no considerable effect in enhancing carrier mobility and only a large shift of threshold voltage ~25 V is observed (See curves 1 and 2 in Figure S3).

On the other hand, the annealing experiment provides a chance to roughly estimate the order of impurity density. In the device, the shift of 25 V corresponds to a density variation of ~2×10$^{12}$ cm$^{-2}$. We note that the real density could be even higher than 2×10$^{12}$ cm$^{-2}$ because a complete desorption under vacuum of $10^{-4}$ Pa is difficult from the experience in surface science. A complete gas desorption requires an ultrahigh vacuum (UHV) surrounding which is unavailable in our lab. In addition, chemical residues likely stay on the surface even after UHV annealing. To remove them on $MoS_2$ is proven harder than on graphene because $MoS_2$ is not as stable as graphene and tends to decompose at elevated temperatures. An effective cleaning process for $MoS_2$ deserves a further investigation.

Finally, we point out that the impurity density is not necessarily a static quantity and is subject to change during measurement. This may be contributed to the re-absorption of gas molecules on top surface and/or charge injection and trapping in the dielectric layer from bottom gate. Comparing curves 2 and 3, a ~10 V restoration of threshold voltage is observed after 26 hour measurement at room temperature. This partially accounts for the variation of our experimental data and is the reason why we use multiple $n_t/n_b$ ratios to fit all data.

4. Effect of $V_g$ on $\mu$

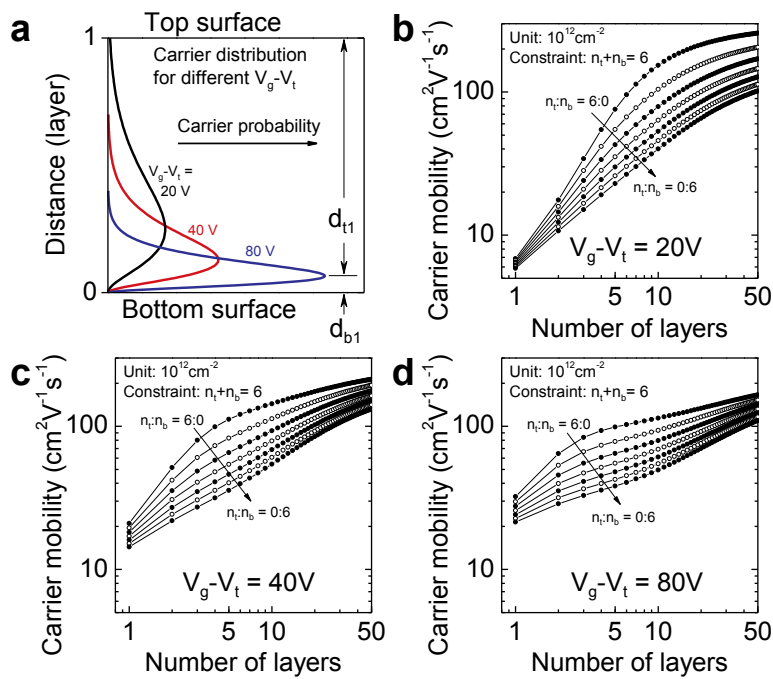

**Figure S4** The effect of carrier density ($\propto V_g - V_t$) on $\mu$. (a) The carrier distribution at different $V_g - V_t$ values of 20, 40 and 80 V. At high $V_g - V_t$, the carriers are close to gated dielectric but the screening of exteneral CI potential is enhanced. (b)–(d) Calculated $\mu$ at three typial $V_g - V_t$ values.

5. Global fit with SR scattering

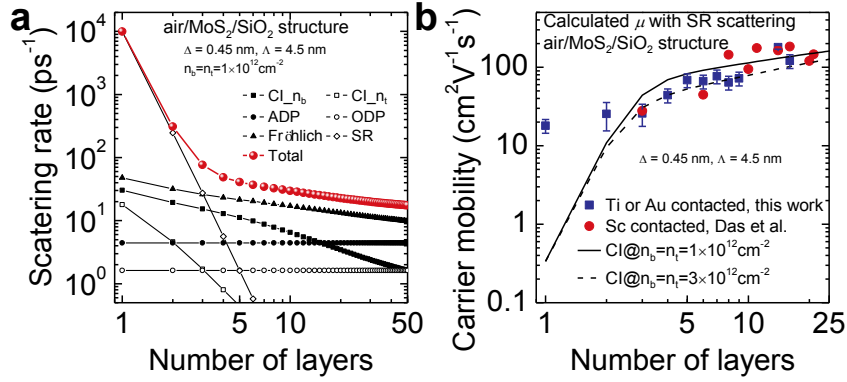

**Figure S5** Calculated scattering rate ($\tau$) and carrier mobility ($\mu$) including SR scattering with roughness parameters $\Delta = 0.45$ nm and $\Lambda = 4.5$ nm. (a) Calculated $\tau$ values for an air/MoS$_2$/SiO$_2$ structure. (b) Comparison between experiment and calculation.

In Si FETs or superlattices, the channel/dielectric surfaces are inherently rough due to the processing: oxygen implantation or epitaxial growth. In Prange and Nee's model,[S2,S3] the SR scattering is characterized by a thickness variation $\Delta$ and an in-plane correlation length $\Lambda$. Within such a model, the scattering matrix element for SR scattering is given by $U_{SR}(q) = \dfrac{\pi^{5/2}\hbar^2 \Delta\Lambda e^{-(q\Lambda)^2/8}}{m^* t^3}$, where $\hbar$ is the reduced Planck constant and $m^* = 0.52 m_0$ the an electron effective mass. A fixed $m^*$ is used for all $t$ values in calculation since it is insensitive to $t$ in MoS$_2$.[S4]

For exfoliated graphene, the ubiquitous surface ripples are argued to possibly result in strong scattering.[S5] Here, the possible scattering from ripples in MoS$_2$ is treated as surface roughness because they share a similar origin (that is, the potential fluctuation at the channel/dielectric interface). Together with phonon and CI contributions, the SR scattering is plotted in Figure 5a. The parameters of channel roughness ($\Delta = 0.45$ nm and $\Lambda = 4.5$ nm) is adopted with reference to the AFM data from graphene on BN ($\Delta \sim 0.45$ nm)[S6] and suspended MoS$_2$ ($\Lambda/\Delta \sim 10$).[S7] Two CI density sets with normal and low levels of $3\times10^{12}$ and $1\times10^{12}$ cm$^{-2}$ are tried (Figure S5b). Since the SR mechanism results in strong thickness dependent mobility $\mu_{SR} \sim \tau_{SR}^{-1} \sim t^6$, the calculations failed to fit the global $\mu-t$ behavior. In view of this, we do not think ripples are a main scattering mechanism and deserve consideration.

**Reference**

S1. Kumagai, M.; Takagahara, T., Excitonic and Nonlinear-optical Properties of Dielectric Quantum-well Structures. *Phys. Rev. B* **1989**, *40*, 12359–12381.

S2. Prange, R. E.; Nee, T.-W., Quantum Spectroscopy of the Low-Field Oscillations in the Surface Impedance. *Phys. Rev.* **1968**, *168*, 779–786.

S3. Sakaki, H.; Noda, T.; Hirakawa, K.; Tanaka, M.; Matsusue, T., Interface Roughness Scattering in GaAs/AlAs Quantum Wells. *Appl. Phys. Lett.* **1987**, *51*, 1934–1936.

S4. Yun, W. S.; Han, S. W.; Hong, S. C.; Kim, I. G.; Lee, J. D., Thickness and Strain Effects on Electronic Structures of Transition Metal Dichalcogenides: 2H-$MX_2$ Semiconductors (M = Mo, W; X = S, Se, Te). *Phys. Rev. B* **2012**, *85*, 033305.

S5. Katsnelson, M. I.; Geim, A. K., Electron scattering on microscopic corrugations in graphene. *Phil. Trans. R. Soc. A* **2008**, *366*, 195–204.

S6. Xue, J.; Sanchez-Yamagishi, J.; Bulmash, D.; Jacquod, P.; Deshpande, A.; Watanabe, K.; Taniguchi, T.; Jarillo-Herrero, P.; Leroy, B. J., Scanning Tunnelling Microscopy and Spectroscopy of Ultra-Flat Graphene on Hexagonal Boron Nitride. *Nat. Mater.* **2011**, *10*, 282–285.

S7. Brivio, J.; Alexander, D. T. L.; Kis, A., Ripples and Layers in Ultrathin $MoS_2$ Membranes. *Nano Lett.* **2011**, *11*, 5148–5153.